\documentclass[fleqn,usenatbib]{mnras}

\usepackage{newtxtext,newtxmath}

\usepackage[T1]{fontenc}
\usepackage{hyperref}

\DeclareRobustCommand{\VAN}[3]{#2}
\let\VANthebibliography\thebibliography
\def\thebibliography{\DeclareRobustCommand{\VAN}[3]{##3}\VANthebibliography}

\usepackage{graphicx}	
\usepackage{amsmath}	
\usepackage{float}
\usepackage{booktabs}

\newcommand{\Msun}{\,{\rm M_\odot}}

\newcommand{\dist}{\, {\rm ckpc} \, h^{-1}}

\def\hmpc{h^{-1}{\rm Mpc}}
\def\hmsun{{h^{-1} \rm M_{\odot}}}
\def\msun{\rm \, M_{\odot}}
\def\hkpc{h^{-1}\, {\rm kpc}}
\def\astrid{\texttt{ASTRID} }
\def\Msigma{$M_{BH} - \sigma$ }
\def\MMstar{$M_{BH} - M_\star$ }
\def\KH{Kormendy \& Ho relation }
\def\calM{{\cal M} }

\usepackage{xcolor}

\title[Over-massive BHs in ASTRID and Illustris TNG50]{Over-massive Central Black Holes in the Cosmological Simulations ASTRID and Illustris TNG50}

\author[E. Weller et al.]{Emma Jane Weller$^{1}$\thanks{emmaweller@college.harvard.edu},
Fabio Pacucci$^{1,2}$\thanks{fabio.pacucci@cfa.harvard.edu}, Priyamvada Natarajan$^{2,3,4}$, Tiziana Di Matteo$^{5,6}$ \\
$^{1}$Center for Astrophysics $\vert$ Harvard \& Smithsonian, Cambridge, MA 02138, USA\\
$^{2}$Black Hole Initiative, Harvard University,
Cambridge, MA 02138, USA\\
$^{3}$Department of Astronomy, Yale University, New Haven, CT 06511, USA\\
$^{4}$Department of Physics, Yale University, New Haven, CT 06520, USA\\
$^{5}$McWilliams Center for Cosmology, Department of Physics, Carnegie Mellon University, Pittsburgh, PA 15213\\
$^{6}$NSF AI Planning Institute for Physics of the Future, Carnegie Mellon University, Pittsburgh, PA 15213, USA
}

\date{\today}

\pubyear{2022}

\begin{document}
\label{firstpage}
\pagerange{\pageref{firstpage}--\pageref{lastpage}}
\maketitle

\begin{abstract}
Recent dynamical measurements indicate the presence of a central SMBH with mass $\sim3\times10^6\Msun$ in the dwarf galaxy Leo I, placing the system $\sim50$ times above the standard, local \MMstar relation. While a few over-massive central SMBHs are reported in nearby isolated galaxies, this is the first detected in a Milky Way satellite. We used the \astrid and Illustris TNG50 LCDM cosmological simulations to investigate the assembly history of galaxies hosting over-massive SMBHs. We estimate that, at the stellar mass of Leo I, $\sim15\%$ of galaxies above the \MMstar relation lie $>10$ times above it. Leo I-like systems are rare but exist in LCDM simulations: they occur in $\sim0.005\%$ of all over-massive systems. Examining the properties of simulated galaxies harboring over-massive central SMBHs, we find that: (i) stars assemble more slowly in galaxies above the \MMstar relation; (ii) the gas fraction in these galaxies experiences a significantly steeper decline over time; and (iii) $>95\%$ of satellite host galaxies in over-dense regions are located above the \MMstar relation. This suggests that massive satellite infall and consequent tidal stripping in a group/dense environment can drive systems away from the \MMstar relation, causing them to become over-massive. As the merging histories of over-massive and under-massive systems do not differ, we conclude that additional environmental effects, such as being in overdense regions, must play a crucial role. In the high-$z$ Universe, central over-massive SMBHs are a signature of heavy black hole seeds; we demonstrate, in contrast, that low-$z$ over-massive systems result from complex environmental interactions.
\end{abstract}




\section{Introduction} \label{sec:intro}

Most local massive galaxies host a central supermassive black hole (SMBH), whose mass, $M_{BH}$ is correlated with properties of the stellar component of the bulge of the host, namely its stellar mass ($M_\star$) and the central velocity dispersion ($\sigma$). These correlations, referred to as the \MMstar and \Msigma relations, suggest co-evolution and growth in tandem of the stars and the central SMBH \citep{Magorrian_1998, Ferrarese_Merritt_2000, Gebhardt_2000, vdB_2016}. 
The community extensively studied these correlations at low and high redshift (see, e.g., \citealt{Reines_Volonteri_2015, Volonteri_Reines_2016, Volonteri_2022, Zhang_2023}), and also at the low-mass end of the black hole (BH) mass distribution (see, e.g., \citealt{Volonteri_PN2009, Pacucci_2018, Nguyen_2019, Baldassare_2020, Greene_2020_IMBH}), where data are scarce. Observations suggest that the mass of the central SMBH is $\sim10^4$ times smaller than the mass of the stellar bulge of the galaxy \citep{Zhang_2023}. The empirically measured slope, normalization, and dispersion of this relation for isolated galaxies have been used to calibrate BH growth models over cosmic time (see, e.g., \citealt{Natarajan_2014, Ricarte_2018, Pacucci_2020}).

Previously, in a couple of isolated nearby galaxies, namely NGC 4486B, NGC 1277, and the dwarf starburst galaxy Henize 2-10, over-massive SMBHs that lie above the \MMstar relation have been reported \citep{Magorrian_1998,vandenBosch+2012,Reines+2011, Schutte_2022}. Meanwhile, in a handful of cases, there are galaxies that appear to host under-massive central SMBHs \citep{Magorrian_1998,Merritt+2001, Nguyen_2019}. However, correlation measurements are yet to be extended to nearby satellite galaxies in the Local Group. 

A dynamical analysis by \cite{Bustamante-Rosell_2021} recently found that the nearby dwarf spheroidal galaxy Leo I hosts a central SMBH with a mass of $(3.3 \pm 2.0) \times 10^6 \Msun$, remarkably similar to Sgr A*, the SMBH at the center of the Milky Way \citep{EHT_2022}. This conclusion was based on precise integral field spectroscopy measurements, which permitted measurement of a rising stellar velocity towards the galactic center. This rise could not be explained by any other galactic component — the absence of a BH is excluded at $>95 \%$ significance in all of the dynamical models considered in their analysis.

Recently, \cite{Pacucci_2022} suggested that accretion from winds of evolved red giant branch stars could make this putative SMBH visible electromagnetically.
Leo I is only $\sim 255 \, \rm kpc$ away, but the presence of its central SMBH remained undetected because it contains very little gas and exhibits no recent star formation due to ram pressure stripping during its infall towards the Milky Way \citep{Mateo_2008,Ruiz-Lara_2021}. In contrast, the dwarf starburst galaxy Henize 2-10, which is $\sim 9 \, \rm Mpc$ away, has an over-massive central SMBH that is actively accreting, and it has been detected as a compact radio and X-ray source at the dynamical center of the galaxy \citep{Reines+2011, Schutte_2022}.

The mass of the central, putative SMBH in Leo I is surprising, as the stellar mass of the galaxy is $(5.2 \pm 1.2) \times 10^7 \msun$ \citep{Mateo_2008}. This places the system above the standard relations, as noted, by a factor of $\sim 50$. Leo I is also remarkable in that the Bondi radius of its central SMBH is similar in size to the core radius of the galaxy \citep{Pacucci_2022}. The core radius characterizes the size of the central distribution of stars assuming, e.g., an isothermal profile \citep{Mateo_2008}. We do not know of any other galaxy whose entire stellar core lies within the gravitational sphere of influence of its central SMBH. Note that the Bondi radius of Sgr A* is significantly smaller because the SMBH is likely fed by Wolf-Rayet stars, characterized by a much higher wind velocity (see, e.g., \citealt{Ressler_2020}).

It is therefore interesting to investigate how the over-massive central SMBH in Leo I might have assembled in concert with the growth of the stellar component of its host. 
The presence of over-massive BHs at $z \gtrsim 8$, possibly detectable as obese black hole galaxies (OBGs) by the James Webb Space Telescope (JWST), is argued to be a salient test of early BH seeding models \citep{Agarwal+2013, Natarajan_2017, Nakajima_2022, Regan_2023}. The existence of local or low-redshift counterparts of these over-massive early BHs is yet to be confirmed.

In this study, we use the \astrid \citep{Ni_2022, Bird_2022} and TNG50 \citep{Pillepich_2018, Nelson_2018, Marinacci_2018, Springel_2018, Nelson_2019_TNG50} cosmological simulation suites to investigate the properties of galaxies hosting over-massive central SMBHs, with the goal of better understanding how common they are and how their assembly histories differ from galaxies that lie on the \MMstar relation. In \S \ref{sec:simulations}, we provide an overview of the simulations used. In \S \ref{sec:rarity}, we investigate how unusual it is for galaxies in \astrid to host significantly over-massive central SMBHs. In \S \ref{sec:tracks}, we study the evolutionary tracks in the \MMstar space for a few sample galaxies in TNG50. In \S \ref{sec:growth}, we compare the stellar and central SMBH growth histories for a larger population of galaxies in TNG50 that host over-massive and under-massive central SMBHs. We investigate what sets apart the evolution of over-massive systems. In \S \ref{sec:stripping}, we focus on the effect of tidal stripping and discuss how the host's location within a galaxy group vs. the field impacts its inferred assembly history. Finally, in \S \ref{sec:conclusion}, we summarize the key results and discuss the implications of our work for understanding the seeding and growth history of central SMBHs and their host galaxies.

\section{Simulations} \label{sec:simulations}

In this Section, we briefly describe the cosmological simulations used. The interested reader is referred to the original papers for more in-depth descriptions.

We work with two different simulation suites to balance their advantages and limitations. \astrid has a smaller BH seed mass, allowing us to study central SMBHs comparable to the one in Leo I. However, TNG50 has merger trees with which we can easily track a given subhalo through time, a feature that \astrid does not currently provide. Several papers (e.g., \citealt{Habouzit_2021}) have investigated how different cosmological simulations with their unique sub-grid models fare in reproducing the \MMstar relation. In particular, they show that for all the simulations, the shape and normalization of the relation vary. For this reason, analyzing a problem from the perspective of multiple different cosmological simulations can bring more insights than using a single simulation.

\subsection{ASTRID} \label{sec:ASTRID}

\astrid is a cosmological hydrodynamical simulation run from $z=99$ to $z=1.7$, with plans to reach $z=1$, using a new version of the \texttt{MP-Gadget} code. It contains $5500^3$ cold dark matter (DM) particles and an initially equal number of smoothed particle hydrodynamic mass elements in a $250 \hmpc$ side box.

\astrid adopts values for the cosmological parameters from \cite{Planck}, with $\Omega_0=0.3089$, $\Omega_\Lambda=0.6911$, $\Omega_{\rm b}=0.0486$, $\sigma_8=0.82$, $h=0.6774$, $A_s = 2.142 \times 10^{-9}$, and $n_s=0.9667$. The simulation has a gravitational softening length $\epsilon_{\rm g} = 1.5 \hkpc$ for both DM and gas particles and a DM particle mass resolution of $9.6 \times 10^6 \Msun$. In the initial conditions, the gas-particle mass is $1.3 \times 10^6 \Msun$.
Note that a baryon mass resolution of $1.3\times 10^6 \Msun$ implies that the lowest-mass galaxies investigated in this paper are resolved with $\sim 100$ stellar particles. We alert the reader that a galaxy resolved with a low number of particles can be more prone to tidal stripping and disruption, which is one of the effects we investigate in this paper.
A variety of sub-grid models, described in detail in the introductory papers \cite{Ni_2022} and \cite{Bird_2022}, are used to implement galaxy and SMBH formation.  This includes supernova and active galactic nucleus (AGN) feedback, inhomogeneous hydrogen, helium reionization, and massive neutrinos. 
In particular, galaxies are identified with the SUBFIND algorithm \citep{Springel_2001}, which identifies and labels particle groups that are self-bound, and locally overdense. 
The most massive galaxies in the group are typically referred to as brightest cluster galaxies (BCGs). The gaseous and stellar mass outside the virial radius of BCGs is referred to as intra-cluster light (ICL). The mass in ICL can artificially increase the stellar mass associated with a given sub-halo, although this effect is important only for very massive galaxies (see, e.g., \citealt{Pillepich_2018}). This does not affect the present work, as we are only interested in over-massive systems with a massive BH in a lower-mass galaxy.
 
In \texttt{ASTRID}, SMBHs are represented as particles that can merge, accrete gas, and apply feedback to their baryonic surroundings. A Friends-of-Friends (FOF) group finder is run periodically, and BHs are seeded in haloes with $M_{\rm halo,FOF} > 5 \times 10^9 \hmsun$ and $M_{\rm *,FOF} > 2 \times 10^6 \hmsun$. BH seed masses are chosen stochastically from a power-law probability distribution with index $n = -1$, ranging from  $3 \times 10^{4} \hmsun$ to $3 \times 10^{5} \hmsun$. A Bondi-Hoyle-Lyttleton-like prescription is used to estimate the accretion rate of gas onto the BHs \citep{DSH2005}.

\astrid uses a newly developed sub-grid dynamical friction model \citep{Tremmel2015,Chen2021} for SMBH dynamics. Since BHs are not artificially pinned to the centers of haloes, as are in most other treatments, a population of wandering, non-central BHs is produced \citep{Weller_2022_ASRID, DiMatteo_2022}, as also noted in the Romulus suite of simulations \citep{Ricarte+2021}. Two BHs merge if their separation is within two times the spatial resolution ($2 \epsilon_g$) after their kinetic energy is dissipated by dynamical friction and they become gravitationally bound to each other.

\subsection{Illustris TNG50} \label{sec:TNG50}

IllustrisTNG \citep{Pillepich_methods, Pillepich_2018, Nelson_2018, Naiman_2018, Marinacci_2018, Springel_2018, Nelson_2019_Illustris} is a suite of cosmological hydrodynamical simulations run from $z=127$ to $z=0$ with the moving-mesh code \textsc{Arepo} \citep{Springel_2010}. It consists of three simulation volumes: TNG50, TNG100, and TNG300. The simulations use the cosmological parameters from \cite{Planck_2016}: $\Omega_{\rm \Lambda,0} = 0.6911$, $\Omega_{\rm m,0} = 0.3089$, $\Omega_{\rm b,0} = 0.0486$, $\sigma_8 = 0.8159$, $h = 0.6774$, and $n_s = 0.9667$.

For our study, we used TNG50-1 \citep{Nelson_2019_TNG50, Pillepich_2019}, the highest resolution run of the TNG50 volume. TNG50 has $2 \times 2160^3$ resolution elements in a box with a side length of $35 \hmpc$. At $z=0$, the collisionless components have a gravitational softening length of $288 \, \rm pc$. The mean baryon and dark matter particle mass resolutions are $8.5 \times 10^4 \Msun$ and $4.5 \times 10^5 \Msun$, respectively.

FOF groups are identified during the simulation, and a SMBH with mass $8 \times 10^5 \hmsun$ is seeded whenever a FOF halo exceeds a mass of $5 \times 10^{10} \hmsun$ and does not yet contain a SMBH. The SMBHs accrete according to an Eddington-limited Bondi accretion rate. Each SMBH is repositioned to the minimum potential location of its host halo at every global integration timestep, and SMBH binaries are assumed to merge promptly \citep{Weinberger_2018}.

\section{Results} \label{sec:results}

In this Section, we investigate the evolution of galaxies with over-massive central SMBHs and assess how common they are. We then focus on what causes the divergence between over-massive and under-massive systems - slower stellar growth and tidal stripping in large galactic groups.

\subsection{How Common Are Leo I-type Galaxies?} \label{sec:rarity}

\begin{figure}
\includegraphics[width=\columnwidth]{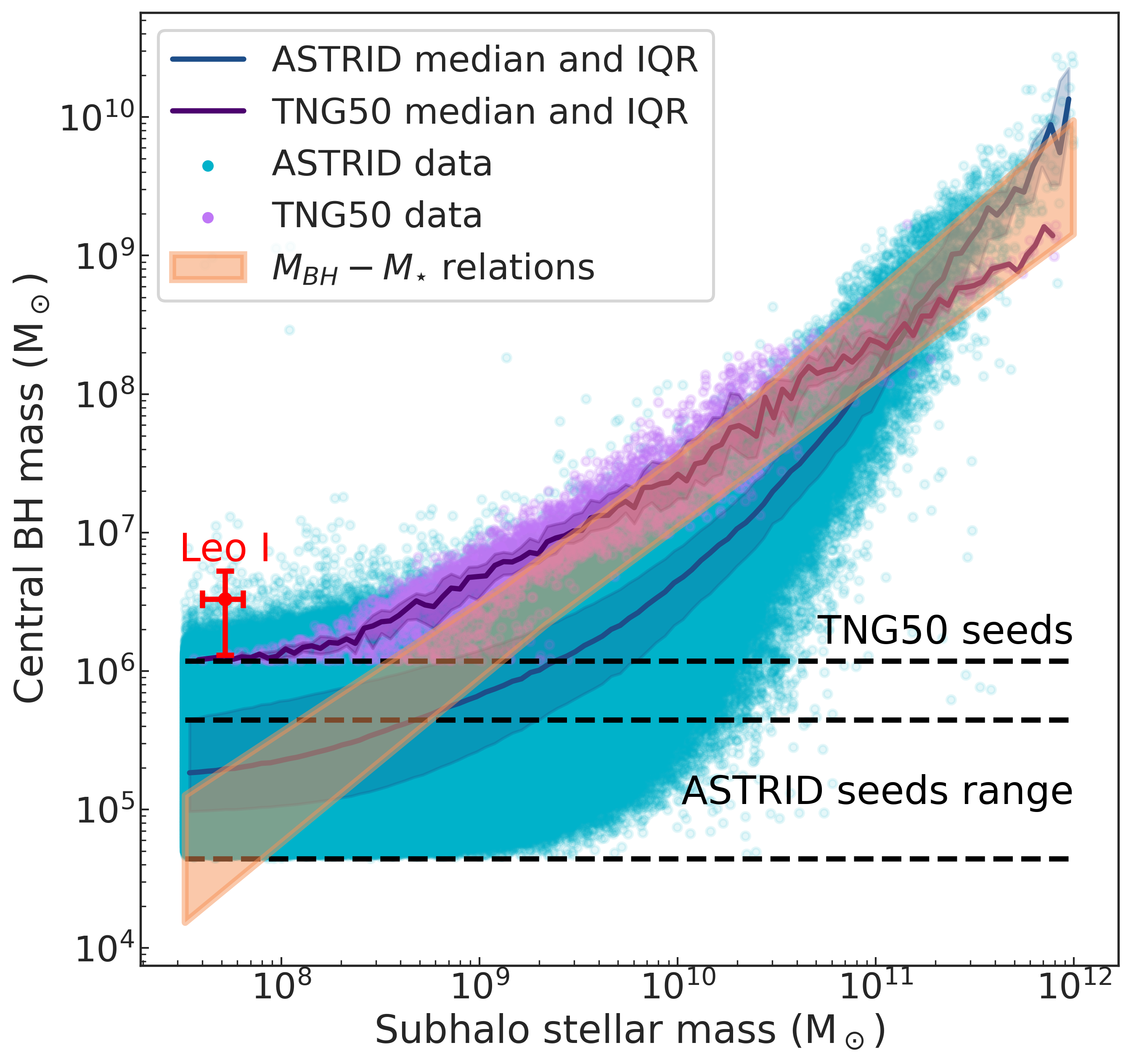}
    \caption{Central BH mass vs. subhalo stellar mass for galaxies in \astrid ($z=2$) and TNG50 ($z=0$), with medians and dispersions. We also mark the location of Leo I with $1\sigma$ uncertainty, the range of empirical $z=0$ \MMstar relations from H{\"a}ring \& Rix (2004), McConnell \& Ma (2013), and Kormendy \& Ho (2013), and the seed masses of each simulation.}
    \label{fig:comb_scatter}
\end{figure}

How unlikely is it for a low-mass galaxy to host a significantly over-massive central SMBH like the one identified in Leo I? In Fig. \ref{fig:comb_scatter}, we display the central BH mass vs. subhalo stellar mass for a large population of galaxies in \astrid and TNG50 — an overview of the galaxies available in the simulations. 
In this population, we considered galaxies with subhalo stellar masses between $\sim 3.3 \times 10^7$ and $10^{12} \Msun$ that contain one or more BHs. Our lower bound was chosen to be the stellar mass at which the \MMstar relation from \cite{Kormendy_Ho_2013} (hereafter referred to as the Kormendy \& Ho relation) crosses \texttt{ASTRID}'s minimum BH seed mass.

In \texttt{ASTRID}, we used all galaxies that meet these criteria at $z=2$ (the lowest-redshift full snapshot available at the time of this writing), for a total of $\sim 3.8 \times 10^6$ galaxies. We considered the most massive BH in each galaxy to be the central BH. In TNG50, we used all $\sim 4.0 \times 10^3$ galaxies that meet the criteria at $z=0$. Each of these galaxies has only one BH: the central one. 

In Fig. \ref{fig:comb_scatter}, we have also marked the median and dispersion (expressed as an interquartile range, IQR) for each simulation. As in \cite{Habouzit_2021}, we have shaded the region encompassing the empirical $z=0$ \MMstar relations from \cite{Haring_Rix_2004}, \cite{McConnell_Ma_2013}, and \cite{Kormendy_Ho_2013}. We see that there are several galaxies in \astrid with similar stellar mass and central BH mass to Leo I and that the data flatten out at low stellar masses, diverging from the \MMstar relation. In this regime, observations are extremely sparse, and the relation is extrapolated from higher masses.

It is apparent that the \astrid and TNG50 data overlap well in the intermediate stellar mass regime. At the low-mass end, discrepancies between \astrid and TNG50 are large. This can be attributed to the fact that in this stellar mass range, the central BH masses are close to the original seed masses, which are significantly different in the two simulations. However, the \MMstar relation is also highly uncertain at such low masses, and could even flatten out altogether (see, e.g., \citealt{Pacucci_2018}). An additional caveat is warranted: although all three empirical \MMstar relations in Fig. \ref{fig:comb_scatter} are based on the observationally determined stellar mass of the bulge, this same property is not easily retrievable from simulations. Instead, we use the total subhalo stellar mass as a proxy. The diversity in galaxy morphologies and properties of stellar haloes can create discrepancies between the model and simulation results, especially at low stellar masses (see, e.g., recent discussions in \citealt{Terrazas2016, Terrazas2017, Habouzit_2021}). Therefore the differences seen in Fig. \ref{fig:comb_scatter} between the \MMstar relation and the simulation medians are at least partially due to the different definitions of stellar mass used in the relation and galaxy populations probed by the significantly different volumes. In addition, comparative studies show that, due to the use of differing sub-grid models, independent cosmological simulation suites yield different shapes and normalization for the \MMstar relation (see the recent study in \citealt{Habouzit_2021}). Although \astrid and TNG50 have somewhat different sub-grid models for galaxy formation, feedback, and initial seeding, we believe that using two different cosmological simulations is helpful for the purposes of our analysis, in order to take advantage of the particular properties of both.

\begin{figure}
	\includegraphics[width=\columnwidth]{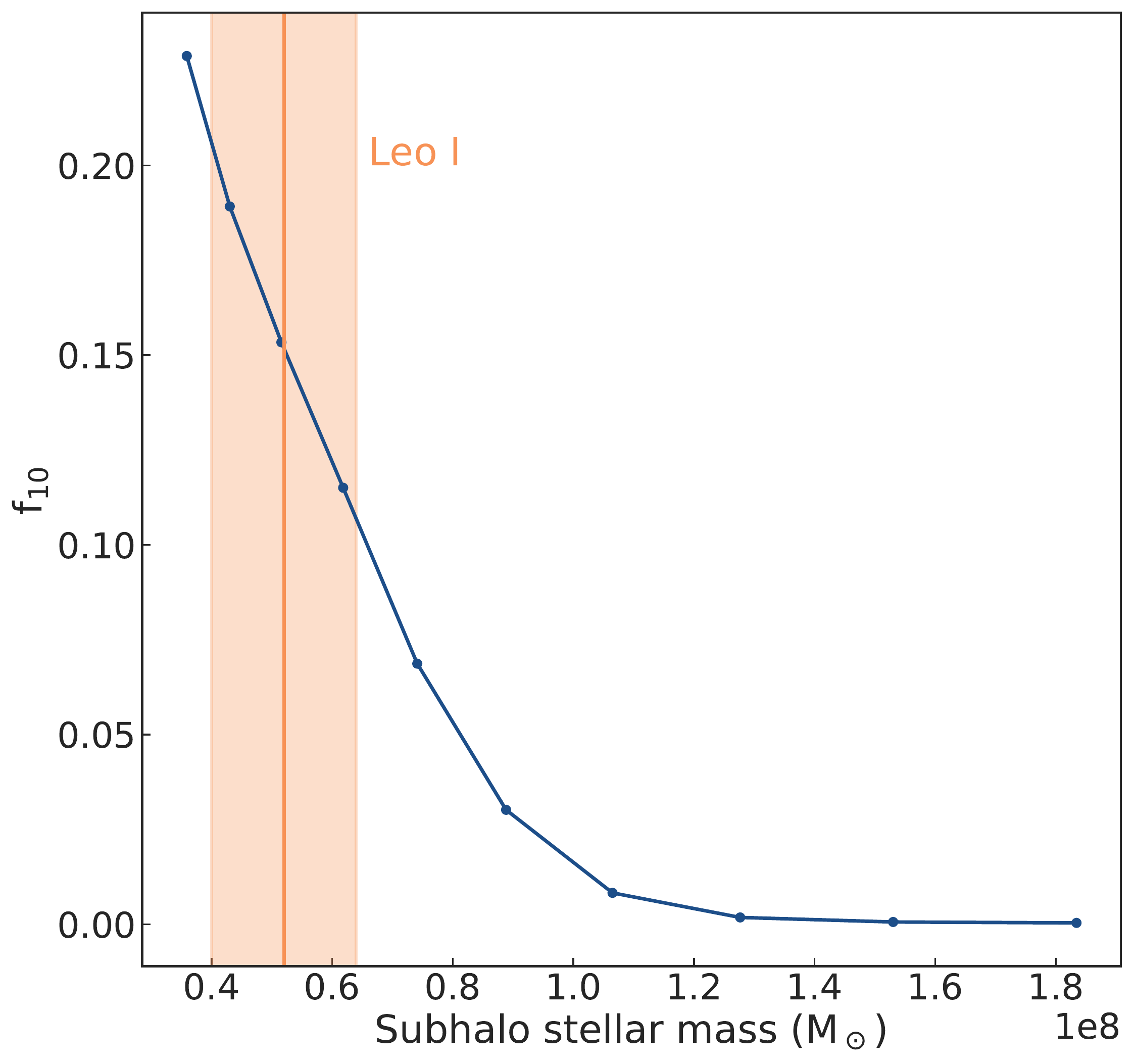}
    \caption{Fraction $f_{10}$ of \astrid galaxies ($z=2$) above the Kormendy \& Ho (2013) \MMstar relation that lie more than 10 times above it. Calculated for ten stellar mass bins ranging from $\sim 3.3 \times 10^7$ to $2 \times 10^8 \msun$. The stellar mass of Leo I and the associated $1\sigma$ error are marked. $f_{10}$ is higher for low stellar masses.}
    \label{fig:frac_above}
\end{figure}

Next, we quantify how rare significantly over-massive central BHs are in our sample of \astrid galaxies. In Fig. \ref{fig:frac_above}, we show the fraction $f_{10}$ of galaxies above the \KH in which the central BH mass is over 10 times larger than the value predicted by the relation. We calculate $f_{10}$ for 10 logarithmically-spaced subhalo stellar mass bins ranging from $\sim 3.3 \times 10^7 \Msun$ to $2 \times 10^8 \Msun$. Above this stellar mass range, $f_{10} \approx 0$.

It is apparent that $f_{10}$ is higher for lower stellar masses. The bin centered at $\sim 5.2 \times 10^7 \msun$, the mass of Leo I, has a fraction $f_{10} \approx 0.15$. Leo I itself lies $\sim 43$ times above the Kormendy \& Ho relation. We calculate $f_{43}$, analogous to $f_{10}$, and find $f_{43} \approx 5.5 \times 10^{-5}$ in the bin centered at Leo I's mass. This conveys that cases like Leo I are rare but exist in \texttt{ASTRID}.

The value of $f_{10}$ expresses the fraction of over-massive central BHs that are significantly above the \MMstar relation. We cannot accurately compute the fraction of all central BHs that are over-massive because, due to the seed mass cutoff, there are very few galaxies below the \MMstar relation at low stellar masses (see Fig. \ref{fig:comb_scatter}). 
We note that BHs are not artificially repositioned in ASTRID, so repositioning is not causing central BHs to be lost. However, there are other ways BHs can be lost, depending on the physical processes that lead to their binary evolution or their inspiral towards the galactic center.
Simulations, including \texttt{ASTRID}, are just starting to account for dynamical friction realistically (see, e.g., \citealt{Tremmel2015, Pfister_2019, Bortolas_2020, Ma_2021}), but these models still rely on significant approximations and sub-grid assumptions, as direct N-body calculations are prohibitive at cosmological scales. Additionally, the process of gravitational recoil is not currently taken into account.

Overall, using only galaxies above this relation in our calculation helps alleviate this problem because we chose our stellar mass range such that the relation is always above the minimum seed mass. However, we note cautiously that this calculation may still be affected by lingering systematic errors.

\subsection{Individual Galaxy Tracks} \label{sec:tracks}
Here, we seek insights into the differences in evolution between galaxies with over-massive central SMBHs and galaxies on the \MMstar relation. To achieve this, we track the growth of central SMBH mass and stellar mass across time for a representative sample of galaxies.

\astrid does not currently have merger trees, which makes it challenging to track the time evolution of galaxies. Hence, we turned to TNG50. However, due to TNG50's relatively high BH seed mass, there are very few low-mass galaxies containing BHs (see Fig. \ref{fig:comb_scatter}). Therefore, here and in \S \ref{sec:growth}, we use galaxies with stellar masses significantly above that of Leo I.

In the following analysis, we chose to look at galaxies with central SMBH masses of $(1 \pm 0.1) \times 10^7 \Msun$ at $z=0$. From this set, we selected for individual study the four galaxies with the lowest stellar mass and the four closest to the \KH at $z=0$.

Note that in TNG50, SMBHs are often ``seeded'' in the galaxies at low redshift instead of the canonical $z\sim 20-30$ (see, e.g., \citealt{BL01, Bromm_Loeb_2003, Lodato_Natarajan_2006}). This is because of the BH seeding procedure described in \S \ref{sec:TNG50}. Thus, we excluded from our analysis galaxies in which the central SMBH has existed in the simulation for less than $8.7 \, \rm Gyr$ (i.e., 55 snapshots).

To quantify how much a galaxy is above or below the \MMstar relation, we define a parameter $\calM$:
\begin{equation} \label{eq:C_defn}
    \calM \equiv \log_{10}{\frac{M_{BH}}{M_{p}}},
\end{equation}
where $M_{BH}$ is the mass of the central SMBH, and $M_p$ is the BH mass predicted by the \KH given the galaxy's stellar mass $M_\star$. In what follows, we label host + BH systems with $\calM>0$ as over-massive and systems with $\calM<0$ as under-massive.

\begin{figure}
    \includegraphics[width=\columnwidth]{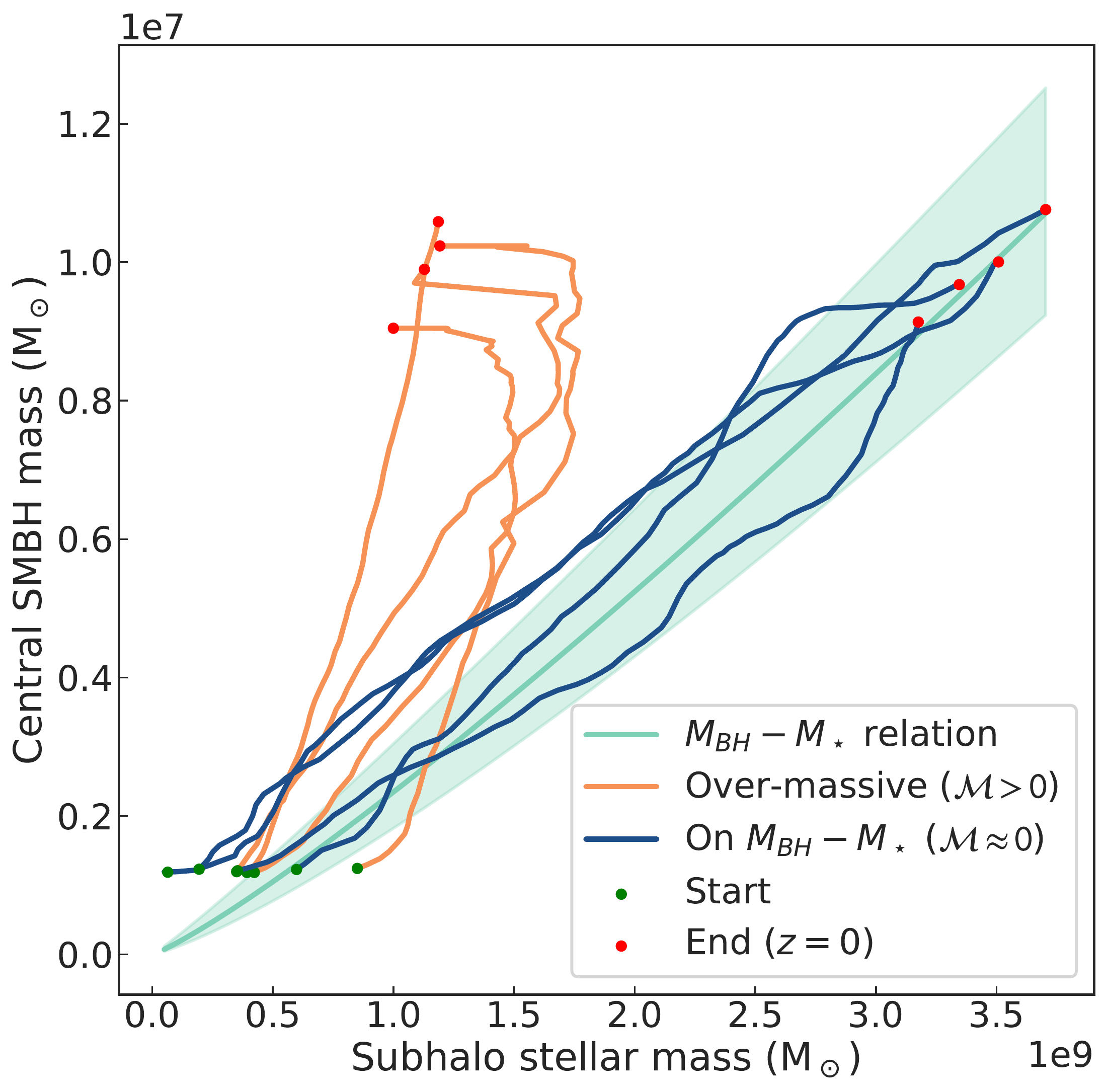}
    \caption{Time evolution of the central SMBH mass vs. subhalo stellar mass for four galaxies in TNG50 with over-massive central SMBHs ($\calM>0$) at $z=0$, and four on the Kormendy \& Ho (2013) \MMstar relation ($\calM \approx 0$) at $z=0$. We see that the $\calM \approx 0$ systems grow along the \MMstar relation, while the $\calM>0$ ones quickly diverge and experience less stellar mass growth. The dispersion in the \MMstar relation is the 1$\sigma$ standard deviation, i.e., the $68\%$ confidence intervals.}
    \label{fig:tracks}
\end{figure}

In Fig. \ref{fig:tracks}, we plot $M_{BH}$ vs. $M_\star$ for the eight galaxies as far back as their central SMBHs exist in the simulation. At $z=0$, the four low-mass galaxies have $\calM>0$, and the four on the \KH have $\calM \approx 0$. Note that the SMBHs appear in the simulation at slightly different redshifts  because of their seeding prescriptions, ranging from $z \approx 1.5$ to $z \approx 2.7$, with a mean of $z \approx 1.6$ for the $\calM>0$ systems and $z \approx 2.2$ for the $\calM=0$ ones. This difference in seeding redshift does not appear to affect the following evolution of the over-massive and under-massive systems.

From Fig. \ref{fig:tracks}, it is evident that the $\calM \approx 0$ systems grow along the \MMstar relation for the entire time period, while the $\calM>0$ ones quickly diverge from the relation, with significantly less growth in stellar mass. We note that this divergence is not a part of our selection criteria but rather a result of it. In three of the four $\calM>0$ systems, the stellar mass even decreases near the end. BH mass growth rates appear slightly higher for $\calM>0$ systems, with a mean of $9.1 \times 10^{-4} \rm \Msun \, yr^{-1}$, versus $8.1 \times 10^{-4} \rm \Msun \, yr^{-1}$ for $\calM \approx 0$ systems.

\subsection{Stellar and Central SMBH Growth} \label{sec:growth}

What drives the cosmological growth of the stars and central SMBHs in the galaxies studied? We now build on the results from \S \ref{sec:tracks} by comparing the growth histories of stellar and central SMBH mass for a larger population of over-massive and under-massive systems in TNG50.

We studied galaxies with stellar masses between $10^9$ and $10^{10} \Msun$ at $z=0$. In this stellar mass range, there are more galaxies above the \KH than below it. Hence, for this section only we adapt the definition of the parameter $\calM$ from \S \ref{sec:tracks} by replacing the \KH with the best-fit line to $\log_{10}{M_{BH}}$ vs. $\log_{10}{M_\star}$ for all galaxies in the chosen stellar mass range. The value of $M_p$ in Eq. \ref{eq:C_defn} is determined by this best-fit line.

\begin{figure}
    \includegraphics[width=\columnwidth]{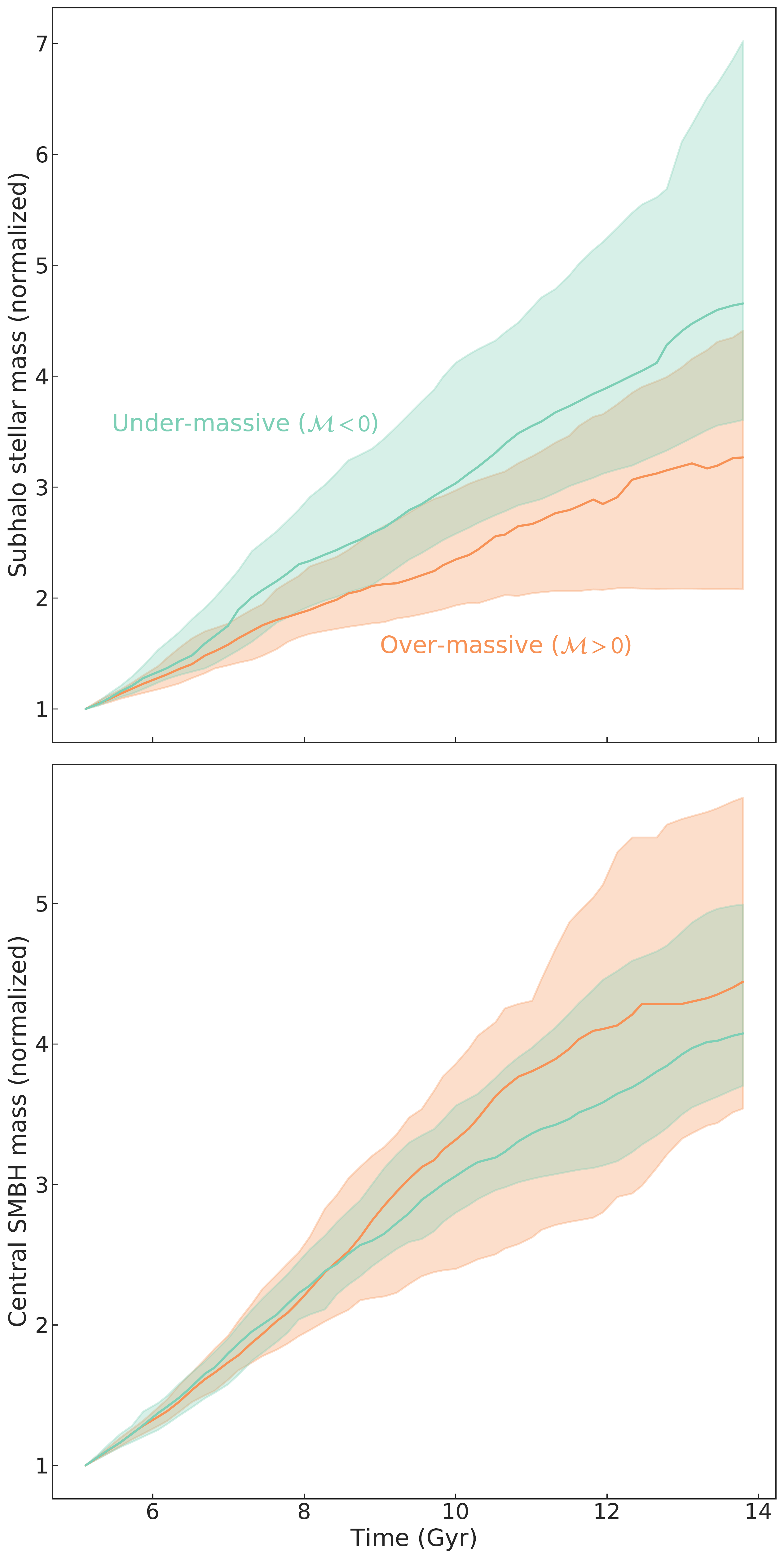}
    \caption{Median and dispersion (IQR) of subhalo stellar mass \textbf{(top)} and central SMBH mass \textbf{(bottom)}, shown over $8.7 \, \rm Gyr$, for systems in TNG50 that are over-massive ($\calM>0$) and under-massive ($\calM<0$) at $z=0$. The masses for each galaxy are normalized to their respective initial values prior to calculating the median and dispersion. We see that in over-massive systems, the stellar mass grows significantly more slowly, while the central SMBH grows marginally more quickly.}
    \label{fig:both_medians}
\end{figure}

We randomly sampled 126 of the 1885 galaxies with BHs in the stellar mass range $10^9$ to $10^{10} \Msun$, and studied these sub-halos for $8.7 \, \rm Gyr$. We excluded any galaxies in which the central SMBH did not exist in the simulation for this whole time period, resulting in a sample of 81 galaxies. Random sampling was performed in order to keep the computational requirements at a reasonable level. We divided the population of galaxies into two groups: 53 with $\calM>0$ at $z=0$ (over-massive), and 28 with $\calM<0$ at $z=0$ (under-massive). In Table \ref{tab:medians}, we list the median subhalo stellar mass and central SMBH mass for each group at the beginning and end of the studied time period. Next, we normalized the stellar and central SMBH mass history of each galaxy to its initial value. In Fig. \ref{fig:both_medians}, we show the median and dispersion (as an IQR) of the galaxies' normalized stellar and central SMBH mass at each snapshot.

\begin{table*}
\centering
\caption{Median subhalo stellar mass, central SMBH mass, and gas fraction for systems in TNG50 that are over-massive ($\calM>0$) and under-massive ($\calM<0$) at $z=0$. Calculated at the beginning and end of the studied $8.7 \, \rm Gyr$ time period, and rounded to 2 significant digits.}
\begin{tabular}{ccccc}
\hline
\textbf{Quantity} & $\mathbf{\calM}$ (at $z=0$) & \textbf{Initial median} & \textbf{Final median} & \textbf{Percent change} \\ \hline
Subhalo stellar mass & $>0$ & $9.6 \times 10^8 \Msun$ & $3.1 \times 10^9 \Msun$ & $220 \%$ \\
 & $<0$ & $6.7 \times 10^8 \Msun$ & $3.3 \times 10^9 \Msun$ & $380 \%$ \\ \hline
Central SMBH mass & $>0$ & $3.2 \times 10^6 \Msun$ & $1.5 \times 10^7 \Msun$ & $360 \%$ \\
 & $<0$ & $2.0 \times 10^6 \Msun$ & $8.4 \times 10^6 \Msun$ & $330 \%$ \\ \hline
 Gas fraction & $>0$ & $0.096$ & $0.058$ & $-40. \%$ \\
  & $<0$ & $0.11$ & $0.084$ & $-20. \%$ \\ \hline
\end{tabular}
\label{tab:medians}
\end{table*}

\paragraph*{Stellar mass growth}
The stellar masses of over-massive and under-massive systems initially grow at similar rates but quickly diverge, with the over-massive systems growing at a slower rate (see also \S \ref{sec:tracks}). One possible explanation for the slower stellar mass growth in over-massive systems is that AGN outflows from the larger central SMBHs suppress and lower the efficiency of star formation \citep{Sturm_2011, Zubovas_King_2016, Zubovas_King_2019}. In particular, the typical energy released by a central SMBH is over two orders of magnitude larger than the binding energy of a host bulge when the system is on the \MMstar relation \citep{King_2015_review}. If the SMBH is over-massive, the energy it releases is even more dominant. This effect is further increased if the over-massive system is a low-mass galaxy because it will have a shallower gravitational potential.

\begin{figure}
	\includegraphics[width=\columnwidth]{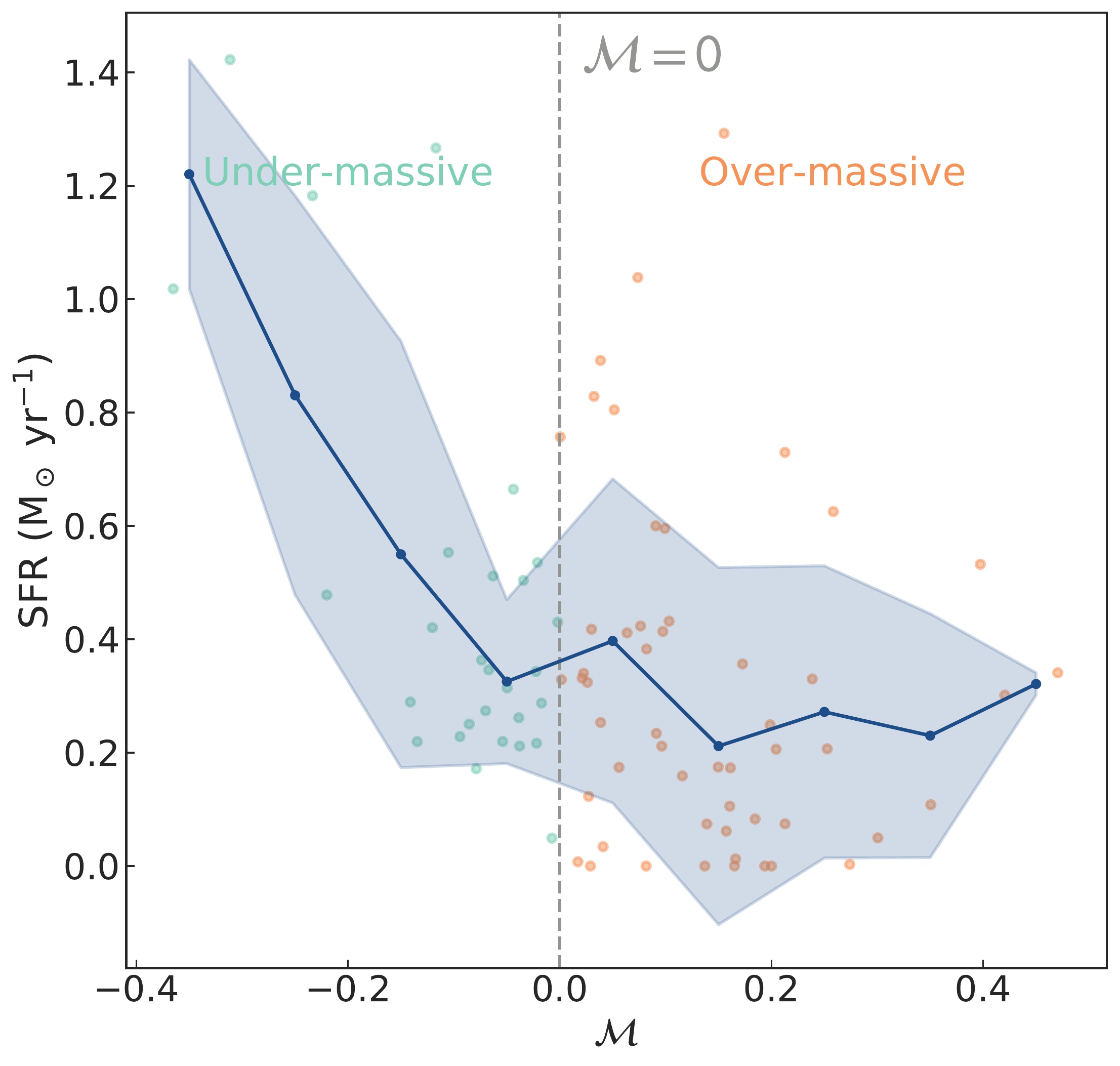}
    \caption{SFR at $z=0$ of galaxies in TNG50 as a function of the $\calM$ value at $z=0$. The mean and standard deviation are also shown in blue for each bin of width $\Delta \calM = 0.1$. The SFR tends to be lower for $\calM>0$ galaxies.}
    \label{fig:both_scatter}
\end{figure}

\begin{figure}
	\includegraphics[width=\columnwidth]{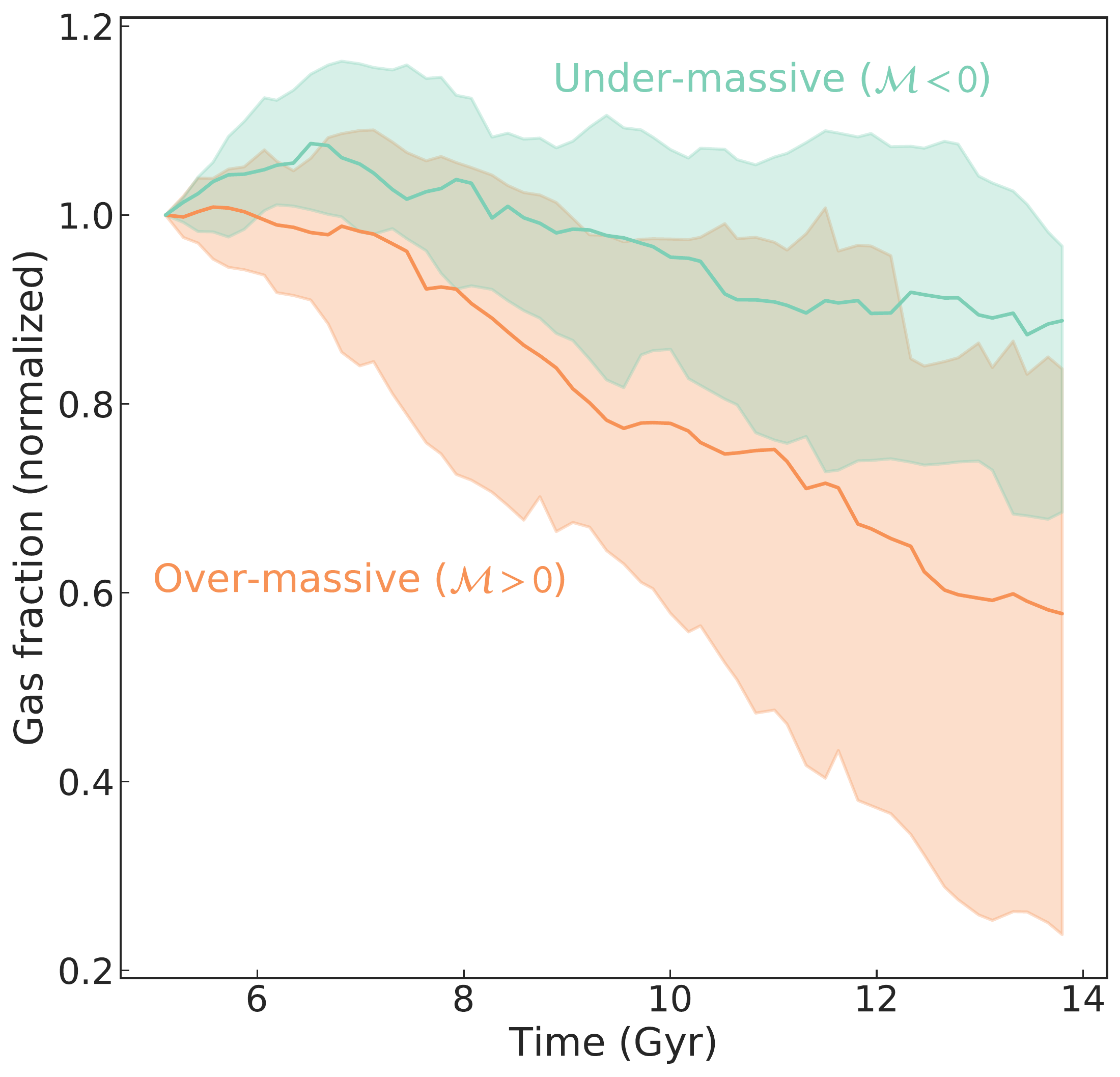}
    \caption{As in Fig. \ref{fig:both_medians}: time evolution of the gas fraction (median and IQR) for over-massive and under-massive systems, normalized to their initial values. The over-massive systems exhibit a significant decline in the gas fraction, as opposed to under-massive systems where the decline is far less pronounced.}
    \label{fig:gasfrac_medians}
\end{figure}

Our findings support this idea. In Fig. \ref{fig:both_scatter}, we display how star formation rate (SFR) correlates with the $\calM$ value at $z=0$ for the 81 galaxies studied. The SFR tends to decrease with increasing $\calM$, so it is typically lower for $\calM>0$ galaxies. In Fig. \ref{fig:gasfrac_medians}, we show that the fraction of gas available to form stars declines more steeply in over-massive systems. As in Fig. \ref{fig:both_medians}, the gas fractions are normalized to the initial value for each galaxy. The initial and final non-normalized medians are given in Table \ref{tab:medians}. The scarcity of gas in over-massive systems is a plausible cause for their lower SFRs. Additionally, the lower SFRs and gas fractions can also be related to tidal stripping (see Sec. \ref{sec:stripping}).

Remarkably, we found no significant difference in the number of major galaxy mergers experienced by over-massive and under-massive systems during the studied time period. We return to this point later in \S \ref{sec:stripping}.

\paragraph*{Central SMBH mass growth}

In the bottom plot of Fig. \ref{fig:both_medians}, we see that the central SMBH mass tends to grow slightly faster for over-massive systems in the last few Gyr, which agrees with the result from \S \ref{sec:tracks}. 

We also investigated whether BH mergers play a significant role in central SMBH growth. We found that only one of our central SMBHs experienced a major merger during the studied period, suggesting that major BH mergers do not drive the growth of the over-massive or under-massive central SMBHs. Instead, these SMBHs grow by steady accretion. Of course, TNG50 only includes BHs in the supermassive regime, so we cannot comment on the number of mergers between the central SMBHs and smaller BHs \citep{Pacucci_2020}.

\subsection{Tidal Stripping} \label{sec:stripping}

The final question we investigate is whether close gravitational interactions with nearby, large galaxies may have caused over-massive systems to diverge from the \MMstar relation via tidal stripping. We calculate what fraction of the stellar component can be removed due to tidal stripping effects. 

\begin{figure}
	\includegraphics[width=\columnwidth]{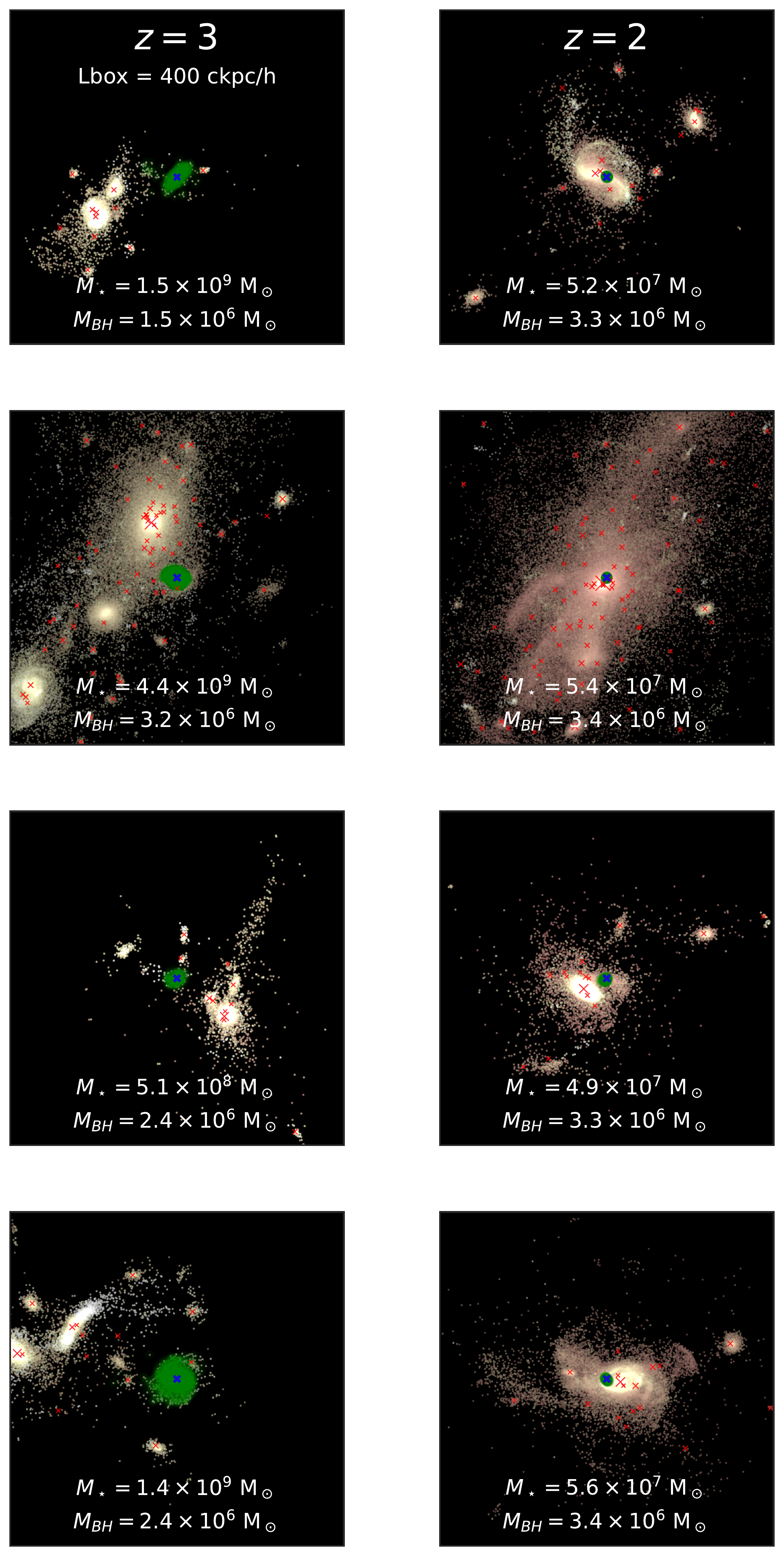}
    \caption{Visualization of four dwarf galaxies in \astrid that at $z=2$ have stellar and central SMBH mass similar to that of Leo I. Shown at $z=3$ (left) and $z=2$ (right). Red crosses represent BHs. The central SMBH is centered and marked with a blue cross, and the stars of its host are highlighted in green. The host galaxies are initially larger but lose most of their stellar mass due to interactions with larger galaxies.}
    \label{fig:galaxy_images}
\end{figure}

First, to obtain some qualitative insights, we used \astrid to investigate four galaxies with stellar and central SMBH mass similar to Leo I at $z=2$. In Fig. \ref{fig:galaxy_images}, we display at $z=3$ and $z=2$ all the stars in a $400 \dist$ side box centered on the central SMBH. At $z=3$, these galaxies contain $\sim 2000-17000$ star particles.

We see that the host galaxies were originally larger; then, they interacted with massive galaxies at the center of groups, losing $\sim 90 - 99 \%$ of their stellar mass between $z=3$ and $z=2$. The bulk of tidal stripping occurs at galactocentric distances comparable to the tidal radius (see, e.g., \citealt{King_1962, Pace_2022}). Meanwhile, their central SMBHs grew only slightly. This suggests that tidal stripping is possibly a cause of stellar mass loss in these over-massive systems, along with the suppressed SFR due to the paucity of gas described in \S \ref{sec:growth} (see, e.g., \citealt{Merritt_1985, Mayer_2006,Read_2006}).
This effect is also consistent with the population of BH pairs in which only one is an AGN, as found in \cite{Chen_2022_dual}; in these systems, the host galaxy of the companion BH is stripped of its gas and stellar component, and, as a result, the BH no longer accretes.

However, we note that while galaxy tidal interactions seem to play an essential role in producing galaxies with over-massive central SMBHs, we found in \S \ref{sec:growth} that there is no apparent difference in the number of galaxy mergers between galaxies with over-massive and under-massive central SMBHs. More than direct mergers, what seems to contribute to the formation of over-massive systems is their local environment. More specifically, what matters is whether the host galaxy is a satellite in an over-dense group environment with a large number of massive neighbors or an isolated field galaxy.

\begin{figure}
	\includegraphics[width=\columnwidth]{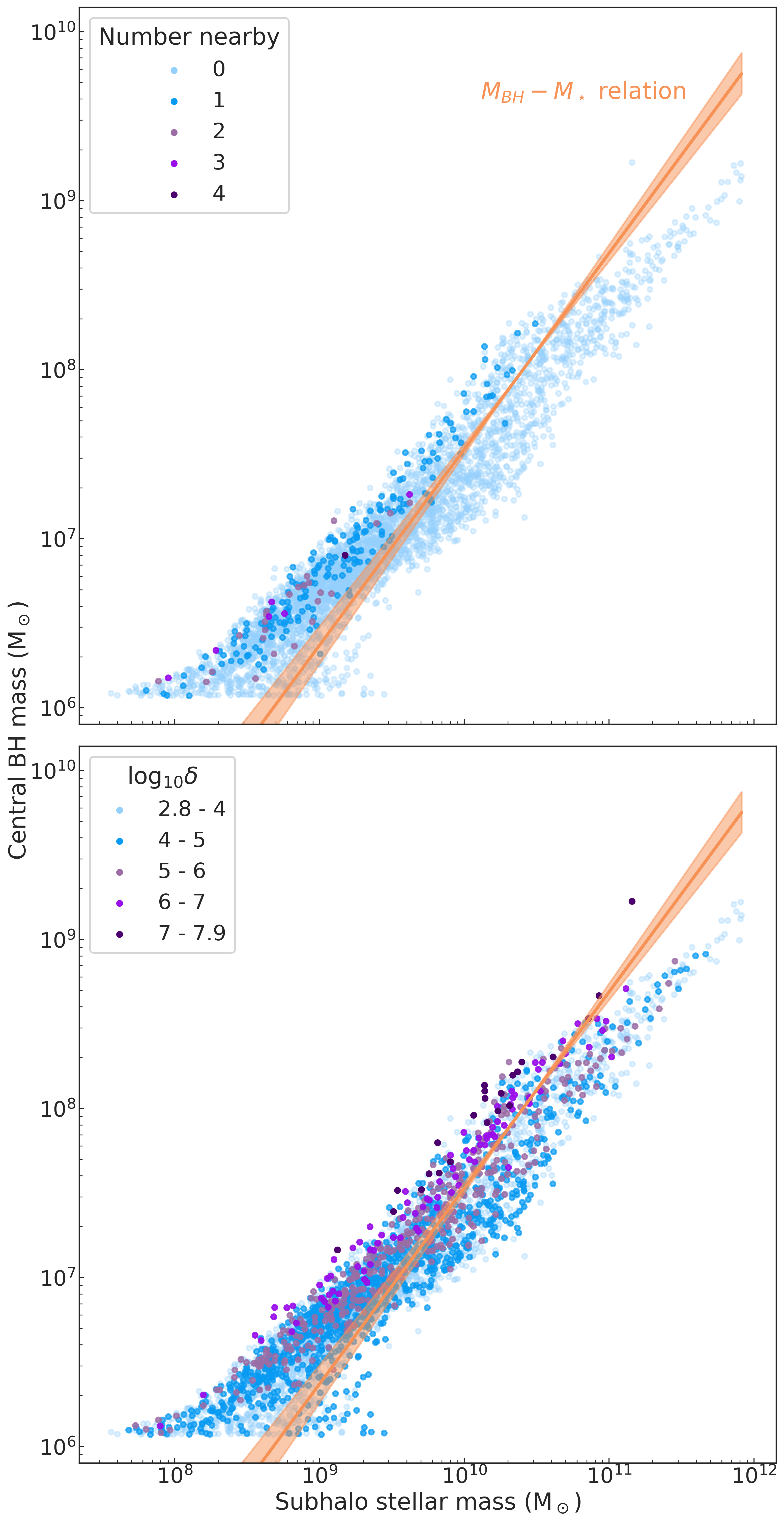}
    \caption{As in Fig. \ref{fig:comb_scatter}, but only including data from TNG50, and only the \MMstar relation from Kormendy \& Ho (2013) is shown. Each galaxy is colored according to: \textbf{Top,} the number of subhalos within $200 \dist$ that have a stellar mass $>50$ times its own; \textbf{Bottom,} the overdensity $\delta$ in which it sits. Nearly all systems that have massive neighbors and/or sit in over-dense environments are over-massive.}
    \label{fig:neighbors}
\end{figure}

In Fig. \ref{fig:neighbors}, we show the TNG50 galaxies from Fig. \ref{fig:comb_scatter}. For each galaxy marked on the top plot, the color indicates the number of subhalos within $200 \dist$ that have a stellar mass over 50 times greater than that of the marked galaxy. We can consider galaxies with one or more of these massive neighbors to be satellite galaxies. We find that nearly all satellite galaxies lie above the Kormendy \& Ho relation. In particular, $\sim 98 \%$ of systems with $\geq 1$ massive neighbors, and $100 \%$ of those with $\geq 2$ massive neighbors, are over-massive. 

In the bottom plot of Fig. \ref{fig:neighbors}, each galaxy is instead colored by the overdensity $\delta$ in which it sits, defined as:
\begin{equation}
    \delta \equiv \frac{\rho - \bar{\rho}}{\bar{\rho}},
\end{equation}
where $\rho$ is the matter density within the half-mass radius of the galaxy, and $\bar{\rho} = \Omega_{m,0} \times \rho_{c,0}$, with $\rho_{c,0}$ the critical density of the Universe (see, e.g., \citealt{Hogg_2003}). 

This plot shows that most galaxies in over-dense environments lie above the Kormendy \& Ho relation. We calculate that $\sim 90 \%$ of systems with $\delta > 10^6$, and $100 \%$ of those with $\delta > 10^7$, are over-massive. We also find that all systems that have both $\geq 1$ massive neighbor and $\delta > 10^6$ are over-massive. 

Overall, Fig. \ref{fig:neighbors} indicates that tidal stripping due to close encounters with neighbors in dense environments drives over-massive systems away from the \MMstar relation. However, the majority of over-massive systems do not belong to groups and do not lie in over-dense environments. Among over-massive systems, only $\sim 7 \%$ have $\geq 1$ massive neighbors, and just $\sim 4 \%$ have $\delta > 10^6$. This suggests that many systems become over-massive entirely via other mechanisms, such as slower star formation (see \S \ref{sec:growth}).

Tidal stripping can explain why three of the four $\calM>0$ systems in Fig. \ref{fig:tracks} experienced a decrease in stellar mass at $z \sim 0$. It is also a possible explanation for the over-massive central SMBH in Leo I, which is a satellite galaxy of the Milky Way \citep{Ruiz-Lara_2021, Pace_2022} and is likely to have been dynamically modified due to tidal interactions. The brightness profile of Leo I does not seem to indicate the occurrence of significant tidal stripping, although the presence of tidal tails was already suggested in the literature (see, e.g., \citealt{Lokas_2008}). 

\section{Discussion and Conclusions} \label{sec:conclusion}

In this work, we used the cosmological simulations \astrid and Illustris TNG50 to investigate the formation and evolution of galaxies with over-massive central SMBHs and to estimate how rare they are. Our main results are summarized below:
\begin{itemize}
    \item More galaxies have significantly over-massive central SMBHs at low stellar masses. At the stellar mass of Leo I, $\sim 15 \%$ of galaxies above the \cite{Kormendy_Ho_2013} \MMstar relation lie more than 10 times over it. Over-massive systems like Leo I are even rarer: they occur in only $\sim 0.005 \%$ of the over-massive systems analyzed in \texttt{ASTRID}.
    \item Galaxies with over-massive central SMBHs at $z=0$ tend to have slower stellar mass growth over time than those on or below the relation. 
    \item This decreased stellar mass growth is likely due to feedback from the over-massive central SMBHs, which inhibit star formation, and tidal stripping from interactions with nearby galaxies. 
    \item Over-massive systems experience a significantly steeper decline of the host's gas fraction with time.
    \item Nearly all satellite host galaxies in over-dense regions have over-massive central SMBHs. Every system with $\geq 1$ nearby massive galaxy and an overdensity of $\delta > 10^6$ is over-massive.
    \item Galaxies above and below the \MMstar relation, on average, experience the same number of galaxy mergers.
    \item Over-massive central SMBHs tend to grow marginally more quickly than other central SMBHs. However, the central SMBHs studied do not experience frequent major BH mergers, indicating that their growth is driven primarily by steady accretion. 
    \item Because the galaxy and BH merger histories of over-massive and under-massive systems do not differ, additional environmental effects must be involved, such as being located in an over-dense region.
\end{itemize}

In summary, we have shown that Leo I-like systems are extremely rare, but they do appear in cosmological simulations. Systems with very over-massive SMBHs at their centers typically grow their stellar component in a slower fashion, and/or they lose part of it because of tidal interactions with nearby massive satellite galaxy neighbors in their group environment. 

It is beyond the scope of this paper to perform a detailed analysis of BH-galaxy relations in the simulations. Careful studies have recently been carried out, for example, by \cite{Habouzit_2021}. We compare to empirical relations to show that simulations do produce a small population of over-massive BHs at low masses. Gaining insight into the environmental properties of systems similar to Leo I may also benefit future follow-up work, which should include higher-resolution zoom-in simulations.

It is also important to note that in this study, we have not considered any difference in the initial seeding mechanisms — over-massive systems are not necessarily seeded by heavier BH seeds at high redshift, as the simulations studied here implement disparate seeding prescriptions. Understanding the origin of these over-massive systems — and possibly detecting more of them at low redshift, where they can be studied in detail — will possibly advance our understanding of the growth history of galaxies and how they were seeded with SMBHs in the first place.

\section*{Acknowledgements}

This work was completed in part as a class project for Astronomy 98: Research Tutorial in Astrophysics, taught at Harvard College by Ramesh Narayan. We thank Vadim Semenov, Lars Hernquist, and the referee for their constructive comments on the manuscript. F.P. acknowledges support from a Clay Fellowship administered by the Smithsonian Astrophysical Observatory. This work was also supported by the Black Hole Initiative at Harvard University, which is funded by grants from the John Templeton Foundation and the Gordon and Betty Moore Foundation.

\section*{Data Availability}
Part of the \astrid data is available at \url{https://astrid-portal.psc.edu/}. The IllustrisTNG data are available at \url{https://www.tng-project.org/}, as described in \cite{Nelson_2019_Illustris}. The codes used to analyze the data will be shared on reasonable request to the corresponding author.



\bibliographystyle{mnras}
\bibliography{ms} 




\bsp	
\label{lastpage}
\end{document}